\documentclass[final,authoryear,12pt]{article}
\usepackage{abstract}
\usepackage[margin=2cm]{geometry}
\usepackage{setspace}
\usepackage{times}
\usepackage{afterpage}

\usepackage{amssymb}
\usepackage{wasysym}
\usepackage{setspace} 
\usepackage{savesym}
\savesymbol{iint}
\savesymbol{iiint}
 \usepackage{amsmath}
\usepackage[version=3]{mhchem}
\usepackage{graphicx}
 \usepackage{lineno} %% The lineno packages adds line numbers.
 
\usepackage{natbib} % load the package
\usepackage{longtable}

\title{\Large A secular increase in continental crust nitrogen during the Precambrian}
\author{Benjamin W Johnson and Colin Goldblatt} 
\date{}
\begin{document}
\maketitle

\begin{abstract}
Recent work indicates the presence of substantial geologic nitrogen reservoirs in the mantle and continental crust. Importantly, this geologic nitrogen has exchanged between the atmosphere and the solid Earth over time. Changes in atmospheric nitrogen (i.e., atmospheric mass) have direct effects on climate and biological productivity. It is difficult to constrain, however, the evolution of the major nitrogen reservoirs through time. Here we show a secular increase in continental crust nitrogen through Earth history recorded in glacial tills (2.9 Ga to modern), which act as a proxy for average upper continental crust composition. Archean and earliest Palaeoproterozoic tills contain 66$\pm$100 ppm nitrogen, whereas Neoproterozoic and Phanerozoic tills contain 290$\pm$165 ppm nitrogen, whilst the isotopic composition has remained constant at $\sim$4$\permil$.  Nitrogen has accumulated in the continental crust through time, likely sequestered from the atmosphere via biological fixation. Our findings support dynamic, non-steady state behaviour of nitrogen through time, and are consistent with net transfer of atmospheric N to geologic reservoirs over time. 
\end{abstract}

%\linenumbers
\section*{Introduction}
The evolution of the Earth System N cycle and the distribution of N in the Earth over the planet's history are not well constrained \citep{Zerkle_and_Mikhail_2017}. Nitrogen moves between different reservoirs in the Earth system including the atmosphere, biosphere, and geosphere \citep{Marty_1995, Boyd_2001, Busigny_et_al_2003, Busigny_et_al_2011}. 
Changes in the distribution of N  among the major reservoirs of the Earth (mantle, crust, and atmosphere), have direct effects on planetary habitability. Biologic productivity based on N-fixing can be limited under very low N$_2$ partial pressures \citep{Klinger_et_al_1989}, and the amount and speciation of N in the atmosphere affect  temperature through direct or indirect greenhouse warming \citep{Goldblatt_et_al_2009, Byrne_and_Goldblatt_2015, Wordsworth_and_Pierrehumbert_2013}.

Higher-\ce{N2} atmospheres can enhance the effectiveness of greenhouse gasses \citep{Goldblatt_et_al_2009, Wordsworth_and_Pierrehumbert_2013}, potentially providing a solution to the Faint Young Sun Paradox \citep{Sagan_and_Mullen_1972,Fuelner_2012}. Specifically, pressure-broadening \citep{Goldblatt_et_al_2009} of \ce{CO2} by an atmosphere with 2-3 fold more \ce{N2} can provide warming consistent with constraints on atmospheric \ce{CO2} content in the Archean \citep{Sheldon_2006}. It is difficult to assess this, and other hypotheses of changing atmospheric mass \citep{Som_et_al_2012, Som_et_al_2016, Barry_and_Hilton_2016}, through direct measurements of palaeoatmospheric conditions. Another approach is to constrain the history of geologic N reservoirs. 

One such reservoir is the continental crust. Current estimates for the amount of N  in the modern continental crust range from 0.25 present atmospheric N mass (PAN, or 4$\times$10$^{18}$ kg N) \citep{Rudnick_and_Gao_2014} to 0.5 PAN \citep{Goldblatt_et_al_2009,Johnson_and_Goldblatt_2015}. These estimates rely on measurements of individual rock types, which are then weighted by their proportion in the crust. For comparison, estimates of N in the Earth's interior range from 1 to 7 PAN in the Bulk Silicate Earth and $\>$50 PAN in the core \citep[][and references therein]{Johnson_and_Goldblatt_2015}. Modern subducted N is estimated to be 5$\times10^{-10}$ PAN per year \citep{Johnson_and_Goldblatt_2015} with non-arc outgassing of 1.75$\times10^{-11}$ PAN per year \citep{Cartigny_and_Marty_2013}. The estimates of crustal N content may be biased, though, due to the effects of differential chemical weathering and alteration. In addition, these approaches offer no temporal resolution. As an alternative approach, we present measurements of glacial tills through time as a proxy for the upper continental crust. 

Large glaciers and ice sheets erode a wide variety of rock types, and resulting glacial till will represent an average composition of the crust over which they erode. Thus, integration of many samples of glacial till can act as a proxy for average upper continental crust composition. This approach was first utilized by \citet{Goldschmidt_1933}, but has since been used to estimate the upper continental crust composition of both Phanerozoic, juvenile crust \citep{Canil_and_Lacourse_2011} as well as the composition of the crust through time \citep{Gaschnig_et_al_2016}. Physical weathering and erosion by a glacier should not impart any isotopic fractionation on the samples. In addition, while weathering can produce locally distinct $\delta^{15}$N values \citep{Boyd_2001}, it is expected that large glaciers will represent an average composition, which will integrate local variation. 

While biologic N cycling \citep{Gruber_and_Galloway_2008} has been a topic of research for well over a hundred years \citep{Breneman_1889}, the geologic N cycle and exchange of N between the atmosphere and solid Earth have received far less attention.  Some modelling efforts suggested near steady-state N concentrations in the crust, mantle, and atmosphere over at least the Phanerozoic \citep{Berner_2006}, and possibly for most of Earth history \citep{Zhang_and_Zindler_1993,Tolstikhin_and_Marty_1998}. In contrast,  geochemistry \citep{Mitchell_et_al_2010,Busigny_et_al_2011,Barry_and_Hilton_2016}, other models \citep{Hart_1978, Stueken_et_al_2016}, and physical proxies \citep{Som_et_al_2012,Kavanagh_and_Goldblatt_2015,Som_et_al_2016} directly contradict the steady-state hypothesis. The later proxies are consistent with  movement of N between different reservoirs of the Earth and significant changes in the mass of the atmosphere over time.  Additional thermodynamic calculations argue that the evolution of mantle redox and Eh-pH state at subduction zones directly affects \ce{N2} outgassing, and therefore the distribution of N in the Earth through time \citep{Mikhail_and_Sverjensky_2014}. 

Either the distribution of N among the main reservoirs of the Earth (atmosphere, mantle, continental crust) has been in steady-state over Earth history or it has been more dynamic. A difficulty in assessing the validity of steady-state and dynamic interpretations of N distribution over Earth history is reconstructing N concentrations in geologic reservoirs in the past. The analysis of glacial tills presented herein suggests an increase in continental N through time, providing a temporal constraint on one of the three major N reservoirs of the Earth system.

\section*{Nitrogen in glacial tills}
We analyzed a series of tills from \citet{Gaschnig_et_al_2016} for N concentration and N isotopes. These till samples consisted of predominantly fine-grained matrix material, and come from formations as old as 2.9 billion years old to formations as young as 0.3 billion years old. We have also included a younger till, Till-4, which is a standard provided by the Geological Survey of Canada. 

\begin{figure*}[]
\includegraphics[width=\textwidth]{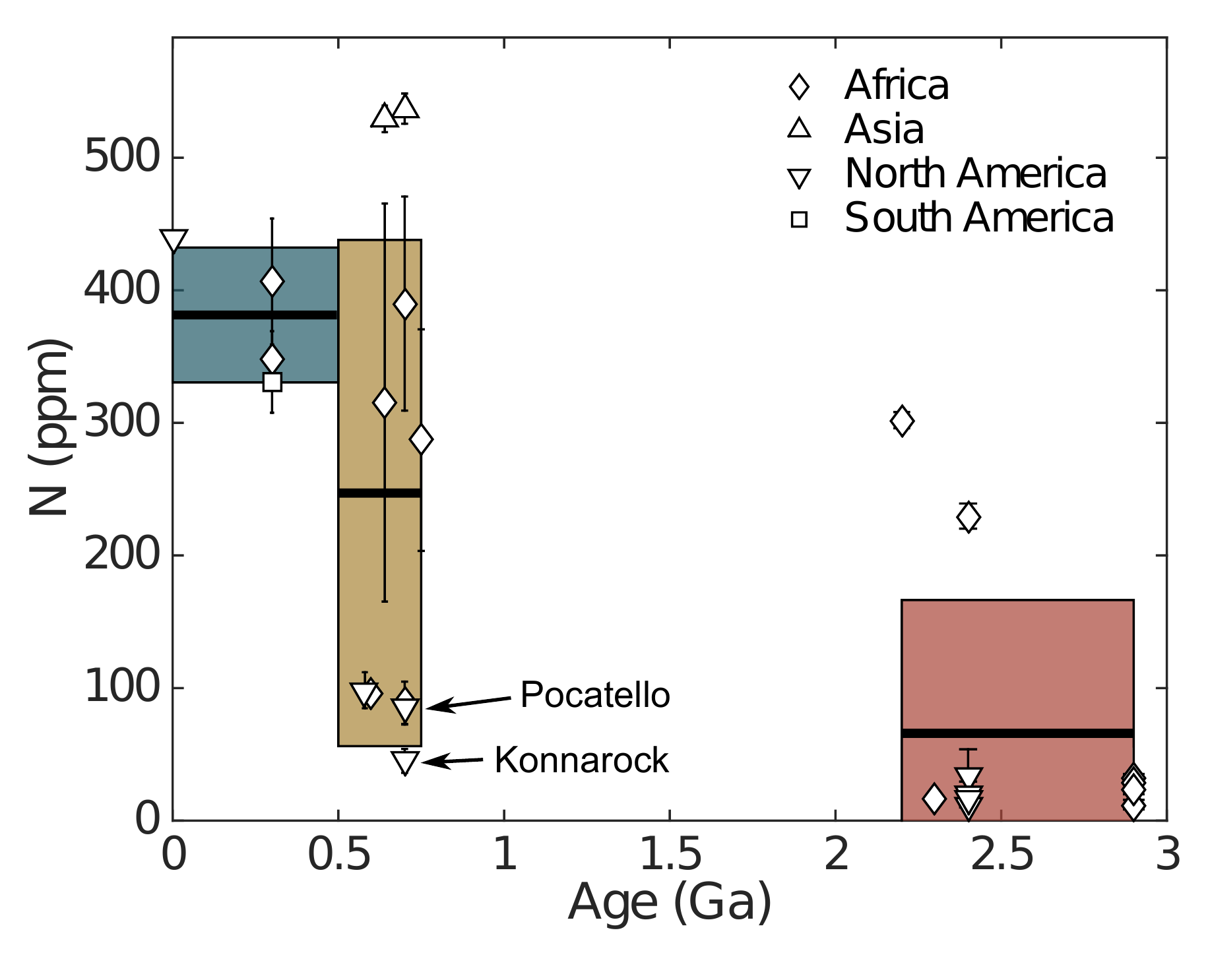}
\caption{Nitrogen concentration in glacial tills through time. Means of triplicate analyses of each sample, with standard deviation, are shown with shapes representing modern continent of exposure. Black lines and coloured boxes show mean and standard deviation of Archean-Palaeoproterozoic, Neoproterozoic, and Phanerozoic samples. Low-N samples from units that have eroded primarily igneous terranes in North America are noted, and discussed in the text.}
\label{fig:Nconc}
\end{figure*}

Nitrogen concentrations are low in glacial tills during the Archean and earliest Palaeoproterozoic, moderate and variable during the Neoproterozoic, and moderate-high and less variable during the Phanerozoic (Fig. \ref{fig:Nconc}, Table \ref{tab:percentages}, Supplemental Information). We define ``low'' as less than average granite, 54 ppm \citep{Johnson_and_Goldblatt_2015}, ``high'' as approaching average upper crust sedimentary rocks, $>400$ ppm, and ``moderate'' as in between. Performing Student's t-test \citep{Student_1908} indicates that both the mean, shown with one standard deviation, Neoproterozoic (250$\pm$180 ppm) and Phanerozoic (380$\pm$50 ppm) concentrations are significantly different from the mean of the Archean and earliest Palaeoproterozoic (66$\pm$100 ppm) samples. There appears to be a secular increase in N content in the continental crust through time. 

\begin{table}[h]
\caption{Proportion of till samples in each age group that have high ($>400$ ppm), low ($<54$ ppm), and moderate (in between) N.}
\begin{center}
\begin{tabular}{l c c c} \hline
\bf{Age} & \bf{$\%$ low}  & \bf{$\%$ moderate}  & \bf{$\%$ high} \\ \hline
Archean & 100 & 0 & 0\\
Palaeoproterozoic & 75 & 25 &  0 \\
Neoproterozoic & 10 & 60 & 30 \\
Phanerozoic & 0 & 50 & 50 \\ \hline
\end{tabular} 
\end{center}
\label{tab:percentages}
\end{table}%

In contrast, mean (plus one standard deviation) $\delta^{15}$N values remain constant within error for all samples, with a value of 3.5$\pm$1.4$\permil$ ~for the Archean and earliest Palaeoproterozoic,  4.9$\pm$4.0$\permil$ ~for Neoproterozoic, and 4.9$\pm$2.6$\permil$ ~ for the Phanerozoic (Fig. $2$). These three populations are not significantly different using Student's t-test. Such isotopic consistency implies either no biologic fractionation during weathering or consistent biologic involvement in glacial weathering through time. 

\begin{figure*}[]
\includegraphics[width=\textwidth]{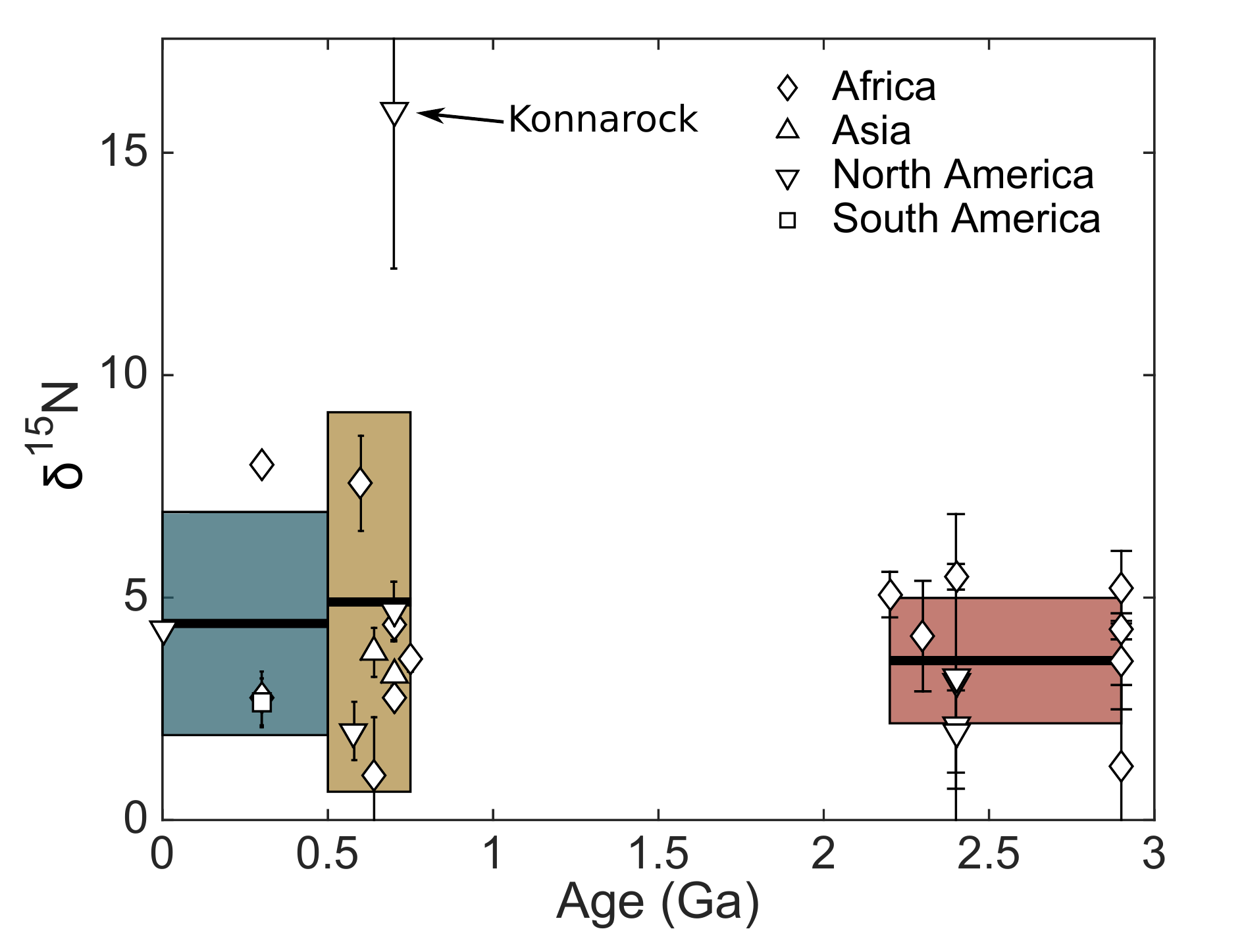}
\caption{Nitrogen isotope values ($\permil$) in glacial tills through time. Averages (black lines) for each time period (Archean-Palaeoproterozoic, Neoproterozoic, and Phanerozoic) are equivalent within error (one standard deviation, coloured boxes), indicating no change in the isotopic character of the continental crust through time. }
\label{fig:Nisotopes}
\end{figure*}

The increase in N concentration through time does not appear to be the result of progressive alteration. There is no correlation between N concentration and $\delta^{15}$N,  the chemical index of alteration (CIA), or Cs/Zr (see Supplemental information). If N was being lost due to weathering or volatilization, low N samples should have high $\delta^{15}$N and CIA values.  If N were behaving as a fluid-mobile element like Cs, there would be a correlation between N and Cs/Zr, with Zr being a non-fluid mobile element.  Such lack of correlation indicates that changes in N concentration are not explained by progressive alteration through time. 

Two of the low-N samples from the Neoproterozoic may not be fully representative of general, contemporaneously formed, upper crust.  One results from erosion of 1.1 Ga Grenville-associated units (Konnarock Formation) and a second is heavily influenced by erosion of bimodal volcanism (Pocatello Formation)  \citep{Gaschnig_et_al_2016}. We suggest Grenvillian rocks may not be representative of the average upper crust, as they typically expose deeper crust from within an orogenic belt. Clasts in the Konnarock Formation are primarily middle to lower crustal granites \citep{Rankin_1993}. Globally, granites average 54 ppm N \citep{Johnson_and_Goldblatt_2015}, much lower than sedimentary or metasedimentary rocks. Though more sparsely measured, volcanic rocks tend to have low N as well, around 0.1 to 10 ppm \citep{Johnson_and_Goldblatt_2015}, owing to the high volatility of N during the eruption of oxidized magma \citep{Libourel_et_al_2003}. Tills that sample only igneous rocks may be biased towards low N.

Additionally, while there is  correlation between N concentration and Rb and K  for low-N samples, there is not for moderate and high-N samples (Supplemental Information). Nitrogen is commonly found as NH$_4^+$ in geologic samples and substitutes for K in silicates \citep{Honma_and_Itihara_1981,Hall_1999}. Many studies have observed correlation between N, K, and Rb  in metasediments \citep{Bebout_and_Fogel_1992, Busigny_et_al_2003}. Thus, low-N samples suggest incorporation of metasedimentary N into the crust via recycling of N into the mantle at subduction zones \citep{Marty_1995, Goldblatt_et_al_2009, Busigny_et_al_2011, Mikhail_and_Sverjensky_2014, Barry_and_Hilton_2016}. Higher N samples imply an additional, or more efficient, transfer mechanism.  

There appears to be a relationship between the present continent of sample outcrop and N concentrations (Fig. \ref{fig:Nconc}). African samples appear to increase in the Palaeoproterozoic and remain high during the Neoproterozoic and Phanerozoic. In contrast, samples from North America are low-moderate into the Neoproterozoic with the most recent sample showing high N concentrations. The single sample from South America and both samples from Asia have moderate to high N. While the strongest control on N concentration appears to be age, it is possible that different continents have a different N history due to differences in their growth history (Supplemental Information). It is also possible that this apparent relationship between present-day geography and N concentration is simply an artifact of a small number of samples. 

\section*{Implications for atmospheric evolution}
How, then, did N accumulate? The isotopic signature is consistent through the record, and is most similar to either modern average marine NO$_3^-$ or sedimentary N (+5 to +7$\permil$). The till record is distinct from both the modern atmosphere (0$\permil$) and the best estimate for the MORB-source mantle value of -5$\permil$, \citep{Marty_1995}. The parsimonious explanation would be incorporation of biologically processed N into the crust with time. Such processing implies N-fixing, thus this N is ultimately of atmospheric origin. 

The mechanisms which transfer N into the continental crust through time are speculative, but have important implications for models of N distribution through time. The most concentrated reservoirs of continental N are in sedimentary and metasedimentary rocks (concentrations 400-500 ppm), with concentrations much higher than igneous rocks  \citep[\textit{e.g.}, 54 ppm in granites, 0.1-10s ppm in basalts, ][]{Johnson_and_Goldblatt_2015}. One likely mechanism of transfer, then, is burial of biologically-processed N at continental margins followed by accretion. An additional mechanism could be input of N at subduction zones. Sparse N concentration and isotopic data  suggests that granitic N content has increased through time (Supplemental Information). In addition, granitic samples show an  increase in $\delta^{15}$N values through time, consistent with enhanced incorporation of biologically processed N. 

The exact timing of incorporation of N is also difficult to determine. There are no glacial deposits from the Mesoproterozoic, rendering this till-based approach ill-suited to this time period. Interestingly, there are two high-N ($>$200 ppm) samples from the Palaeoproterozoic. \citet{Gaschnig_et_al_2016} note a distinct change in the composition of tills between the Archean and Palaeoproterozoic glaciations, reflecting a transition from greenstone/komatiite dominated Archean crust to more felsic crust in the Proterozoic. This trend is perhaps mirrored in some of the N analyses, with more felsic crust having higher N concentration. 

Regardless of the timing of the increase of crustal N, we can compare upper crust N budget from the till proxy to previous work.  \citet{Johnson_and_Goldblatt_2015} suggest $150\pm22$ ppm N in the upper crust, while \citet{Rudnick_and_Gao_2014} suggest 83 ppm. We use a total continental crust mass of $2.28\times10^{22}$ kg \citep{Laske_et_al_2013}, with the upper crust being $53\%$ of the total \citep{Wedepohl_1995}. The \citet{Rudnick_and_Gao_2014} estimate of 83 ppm N  yields $0.1\times10^{18}$ kg N (0.25 PAN) in the upper crust and 150 ppm from \citet{Johnson_and_Goldblatt_2015} suggests 0.5 PAN. Based on exposed and buried outcrop area, the upper continental crust is $28\%$ Phanerozoic,  $31\%$ Neoproterozoic, $16\%$ Mesoproterozoic, $15\%$ Palaeoproterozoic, and $10\%$ Archean \citep{Goodwin_1991, Goodwin_1996}. Given this crust distribution, and assuming the Mesoproterozoic has the same N concentration as the Archean/Palaeoproterozoic, our work suggests an upper crust N concentration of 210 ppm, equivalent to $2.5\times10^{18}$ kg N, or 0.63 PAN (Table \ref{tab:Nbudget}).   
Importantly, the N content of the lower crust is poorly constrained, but could be a significant N reservoir as well. \citet{Johnson_and_Goldblatt_2015} suggest 17 ppm N in the lower crust, which would result in a total continental crust N concentration of 120 ppm and a N mass of $2.7\times10^{18}$ kg.

\begin{table}[h]
\caption{Distribution of upper continental crust ages after \citet{Goodwin_1991, Goodwin_1996}. We assume that tills accurately sample crust of each age, and that the Mesoproterozoic has the same N concentration as the Archean/Palaeoproterozoic.}
\begin{center}
\begin{tabular}{l c c} \hline
\bf{Age} & \bf{$\%$ crust} & \bf{N (ppm)} \\ \hline
Phanerozoic & 28 & 380 \\
Neoproterozoic & 31 & 250 \\
Mesoproterozoic & 16 & 66 \\
Palaeoproterozoic & 15 & 66 \\
Archean  & 10 & 66 \\ 
& \multicolumn{2}{c}{\textit{Total upper crust}} \\
& [N] = 210 ppm & mass = $2.5\times10^{18}$ kg N \\
\hline
\end{tabular}
\end{center}
\label{tab:Nbudget}
\end{table}%

The trend of increased N concentration in the continental crust over time is consistent with non-steady state behavoiur of N through Earth history. Specifically, the till record is consistent with net atmospheric drawdown through time  \citep{Goldblatt_et_al_2009, Busigny_et_al_2011, Barry_and_Hilton_2016}. While the tills provide a constraint on the evolution of one of the three major N reservoirs (continental crust), determining the exact evolution of the other two (mantle and atmosphere) requires more analyses. We cannot necessarily rule out modern or lower \ce{pN2} at specific points in Earth history \citep[e.g.,][]{Som_et_al_2012, Som_et_al_2016,Marty_et_al_2013} but the till data is most consistent with higher atmospheric mass in the past. The balance of mantle outgassing at mid-ocean ridges and arcs to in-gassing at subduction zones is an important, and unconstrained, parameter, over Earth history. Strong net mantle outgassing would be required to have non-decreasing atmospheric N through time.

\section*{Acknowledgments}
The authors thank Richard Gaschnig and Roberta Rudnick for providing glacial till samples as well as Dante Canil for the initial suggestion to use glacial tills as a crust composition proxy. We also thank Natasha Drage at the University of Victoria for assistance with sample compilation. Andy Schauer at the University of Washington assisted in isotopic analyses. Funding was provided in an NSERC Discovery Grant to CG. We thank Sami Mikhail and an anonymous reviewer for constructive feedback, as well as Helen Williams for editorial support. 

\setcounter{figure}{0}   
\setcounter{table}{0}
\renewcommand{\thetable}{S-\arabic{table}}
\renewcommand{\thefigure}{S-\arabic{figure}}

\section*{Supplementary Information}
%\linenumbers
\section*{Sample description and collection}

All samples analyzed were collected by \citet{Gaschnig_et_al_2016}. These samples were collected specifically to assess changes in the composition of the upper continental crust through time. Large ice sheets typically erode a wide variety of rock types, thus till samples should represent an average upper crustal composition. The following is a summary of their collection and sample preparation techniques, but please see the original paper for more detail. 

The sampling strategy focused on collecting fine-grained material. This was achieved by collecting massive diamictite and some drop-stone bearing argillite. In both cases, the fine-grained matrix was crushed in an alumina jaw crusher, clasts larger than 5mm were removed, and the remaining sample crushed to a fine powder using an alumina swing mill. Excepting the Palaeoproterozoic Pecors, Neoproterozoic Blasskranz, and Ordovician Pakhuis formations, all samples are given as composites of each stratigraphic unit. That is, individual crushed samples were homogenized to give a representative average mixture for each formation. 

 \citet{Gaschnig_et_al_2016} determined major element compositions using x-ray fluorescence and trace element composition using laser-ablation ICP-MS techniques. We use their values directly, including chemical index of alteration (CIA), which is calculated as  \ce{Al2O3} / (\ce{Al2O3} + \ce{CaO} + \ce{K2O} + \ce{Na2O}). Note that they corrected CaO to remove any influence of carbonates and apatite.

\section*{Detailed N analytical methods}
All N measurements were done at the University of Washington's IsoLab, following the procedure outlined in  \citet{Stueken_2013}. Briefly, between 10-100 mg of sample powder was weighed into a 9$\times$5 mm Sn capsule. Samples and standards were analyzed on a Thermo-Finnigan MAT 253 coupled to a Costech Elemental Analyzer. Standards used were two glutamic acids (GA-1, GA-2), and two internal standards (dried salmon and organic-rich McRae shale). Samples were flash-combusted at 1000~$^{\circ}$C in a combustion column packed with cobaltous oxide (combustion aid) and silvered cobaltous oxide (sulphur scrubber). Combustion products passed over a reduced copper column at 650~$^{\circ}$C to convert all N to N$_2$ and absorb excess O$_2$. Lastly, sample gas passed through a magnesium perchlorate trap to absorb water and a 3 m gas chromatography column to separate N$_2$ from O$_2$. All analyses were quantified using IsoDat software.  

Errors reported for individual are one standard deviation based on triplicate analysis of each sample. The mean and one standard deviation are shown for each age group (Archean/Palaeoproterozoic, Neoproterozoic, Phanerozoic) are simply calculated from all samples that fall within each age window. Lower N concentrations generally result in greater uncertainty in isotopic measurements due to smaller amounts of N released during analysis. Thus, isotopic uncertainties for the low N samples, most of the  are the Archean/Palaeoproterozoic and some Neoproterozoic, are generally higher (Table \ref{tab:results}). In addition, some error may have been introduced due to not preparing samples in a vacuum. It is possible that some atmospheric \ce{N2} adhered to the powder, though any contamination is suspected to be small due to distinctly non-zero (i.e., non-atmospheric) $\delta^{15}$N values and lack of correlation between N concentration and  $\delta^{15}$N. If atmospheric contamination was a major issue, we would expect for samples with high N to have low $\delta^{15}$N values, which is not observed. This interpretation implies that $\delta^{15}$N values in tills are non-zero initially, which is consistent with observed  $\delta^{15}$N values from a wider variety of continental rocks \citep{Johnson_and_Goldblatt_2015}.

\afterpage{
\begin{longtable}[t]{l c c c c}
\caption{Nitrogen concentration and stable isotopic data. Samples analyzed are from 
\citet{Gaschnig_et_al_2016} and sample names herein are those used in the original publication. 
Age is in Ga, N concentration is in ppm and $\delta^{15}$N is in permil. 
Continent indicates continent where sample was collected: AF - Africa, NA - North America, AS - Asia, SA - South America.}\\

\bf{Stratigraphic unit}	&	\bf{Continent}	&	\bf{Age}	&	\bf{N (ppm)}	&	\bf{$\delta^{15}$N}	\\ \hline
\endfirsthead

\multicolumn{5}{c}%
{{\bfseries \tablename\ \thetable{} -- continued from previous page}} \\ 
\bf{Stratigraphic unit}	&	\bf{Continent}	&	\bf{Age}	&	\bf{N (ppm)}	&	\bf{$\delta^{15}$N}	\\ \hline 
\endhead

\hline \multicolumn{5}{r}{Continued on next page} \\ \hline
\endfoot

\endlastfoot
Mozaan Group	&	AF	&	2.9	&	24	&	4.3	\\
Mozaan Group	&	AF	&	2.9	&	31	&	5.6	\\
Mozaan Group	&	AF	&	2.9	&	28	&	5.8	\\
Afrikander Formation	&	AF	&	2.9	&	9	&	3.1	\\
Afrikander Formation	&	AF	&	2.9	&	11	&	2.8	\\
Afrikander Formation	&	AF	&	2.9	&	16	&	4.8	\\
Promise Formation West Rand Group Witwatersrand	&	AF	&	2.9	&	22	&	4.2	\\
Promise Formation West Rand Group Witwatersrand	&	AF	&	2.9	&	21	&	4.1	\\
Promise Formation West Rand Group Witwatersrand	&	AF	&	2.9	&	28	&	4.5	\\
Coronation Formation West Rand Group Witwatersrand	&	AF	&	2.9	&	33	&	2.0	\\
Coronation Formation West Rand Group Witwatersrand	&	AF	&	2.9	&	35	&	2.5	\\
Coronation Formation West Rand Group Witwatersrand	&	AF	&	2.9	&	30	&	-0.9	\\
Bottle Creek Formation	&	NA	&	2.4	&	10	&	2.8	\\
Bottle Creek Formation	&	NA	&	2.4	&	42	&	0.9	\\
Bottle Creek Formation	&	NA	&	2.4	&	48	&	2.7	\\
Gowganda Formation	&	NA	&	2.4	&	11	&	3.4	\\
Gowganda Formation	&	NA	&	2.4	&	10	&	0.5	\\
Gowganda Formation	&	NA	&	2.4	&	13	&	2.1	\\
Bruce Formation	&	NA	&	2.4	&	19	&	-0.9	\\
Bruce Formation	&	NA	&	2.4	&	12	&	3.6	\\
Bruce Formation	&	NA	&	2.4	&	30	&	6.6	\\
Ramsay Lake Formation	&	NA	&	2.4	&	14	&	3.1	\\
Ramsay Lake Formation	&	NA	&	2.4	&	20	&	3.6	\\
Ramsay Lake Formation	&	NA	&	2.4	&	14	&	3.0	\\
Gowganda Formation	&	NA	&	2.4	&	11	&	3.4	\\
Gowganda Formation	&	NA	&	2.4	&	10	&	0.5	\\
Gowganda Formation	&	NA	&	2.4	&	13	&	2.1	\\
Makganyene Formation	&	AF	&	2.3	&	15	&	4.9	\\
Makganyene Formation	&	AF	&	2.3	&	15	&	4.8	\\
Makganyene Formation Transvaal/Griqualand	&	AF	&	2.3	&	18	&	2.7	\\
Timeball Hill Formation	&	AF	&	2.2	&	295	&	5.2	\\
Timeball Hill Formation	&	AF	&	2.2	&	306	&	4.5	\\
Timeball Hill Formation	&	AF	&	2.2	&	305	&	5.5	\\
Duitschland Formation	&	AF	&	2.4	&	230	&	5.8	\\
Duitschland Formation	&	AF	&	2.4	&	239	&	5.3	\\
Duitschland Formation	&	AF	&	2.4	&	220	&	5.3	\\
Konnarock Formation	&	NA	&	0.7	&	54	&	12.2	\\
Konnarock Formation	&	NA	&	0.7	&	36	&	19.3	\\
Konnarock Formation	&	NA	&	0.7	&	45	&	16.4	\\
Gucheng Formation near bottom of unit	&	AS	&	0.7	&	524	&	3.3	\\
Gucheng Formation	&	AS	&	0.7	&	545	&	3.4	\\
Gucheng Formation	&	AS	&	0.7	&	542	&	3.1	\\
Nantuo Formation lower part of unit	&	AS	&	0.64	&	539	&	4.3	\\
Nantuo Formation middle part of unit	&	AS	&	0.64	&	519	&	3.2	\\
Nantuo Formation top of unit	&	AS	&	0.64	&	530	&	3.8	\\
Pocatello Formation upper diamictite	&	NA	&	0.7	&	99	&	4.6	\\
Pocatello Formation upper diamictite	&	NA	&	0.7	&	78	&	4.1	\\
Pocatello Formation upper diamictite	&	NA	&	0.7	&	78	&	5.4	\\
Blaubeker Formation	&	AF	&	0.7	&	106	&	4.0	\\
Blaubeker Formation	&	AF	&	0.7	&	86	&	4.7	\\
Blaubeker Formation	&	AF	&	0.7	&	74	&	4.4	\\
Kaigas Formation	&	AF	&	0.75	&	315	&	3.8	\\
Kaigas Formation	&	AF	&	0.75	&	353	&	3.7	\\
Kaigas Formation	&	AF	&	0.75	&	193	&	3.3	\\
Numees Formation	&	AF	&	0.6	&	96	&	6.4	\\
Numees Formation	&	AF	&	0.6	&	89	&	7.8	\\
Numees Formation	&	AF	&	0.6	&	100	&	8.5	\\
Ghaub Formation	&	AF	&	0.64	&	163	&	-0.4	\\
Ghaub Formation	&	AF	&	0.64	&	463	&	2.2	\\
Ghaub Formation	&	AF	&	0.64	&	320	&	1.2	\\
Chuos Formation	&	AF	&	0.7	&	306	&	2.8	\\
Chuos Formation	&	AF	&	0.7	&	467	&	3.0	\\
Chuos Formation	&	AF	&	0.7	&	397	&	2.5	\\
Gaskiers Formation	&	NA	&	0.58	&	83	&	1.3	\\
Gaskiers Formation	&	NA	&	0.58	&	109	&	2.6	\\
Gaskiers Formation	&	NA	&	0.58	&	103	&	2.1	\\
Bolivia	&	SA	&	0.3	&	321	&	3.2	\\
Bolivia	&	SA	&	0.3	&	357	&	2.6	\\
Bolivia	&	SA	&	0.3	&	314	&	2.1	\\
DwykaEast Group	&	AF	&	0.3	&	373	&	2.1	\\
DwykaEast Group	&	AF	&	0.3	&	461	&	3.3	\\
DwykaEast Group	&	AF	&	0.3	&	386	&	2.8	\\
DwykaWest	&	AF	&	0.3	&	340	&	8.0	\\
DwykaWest	&	AF	&	0.3	&	372	&	8.0	\\
DwykaWest	&	AF	&	0.3	&	332	&	8.0	\\
Till4	&	NA	&	0.001	&	440	&	4.3	\\
\hline \hline

\label{tab:results}
\end{longtable}%
%\clearpage % omit if you want the material from the main text to continue on this page
} % end of afterpage material

\section*{Constraining post-depositional alteration}

\begin{figure*}[]
\includegraphics[width=\textwidth]{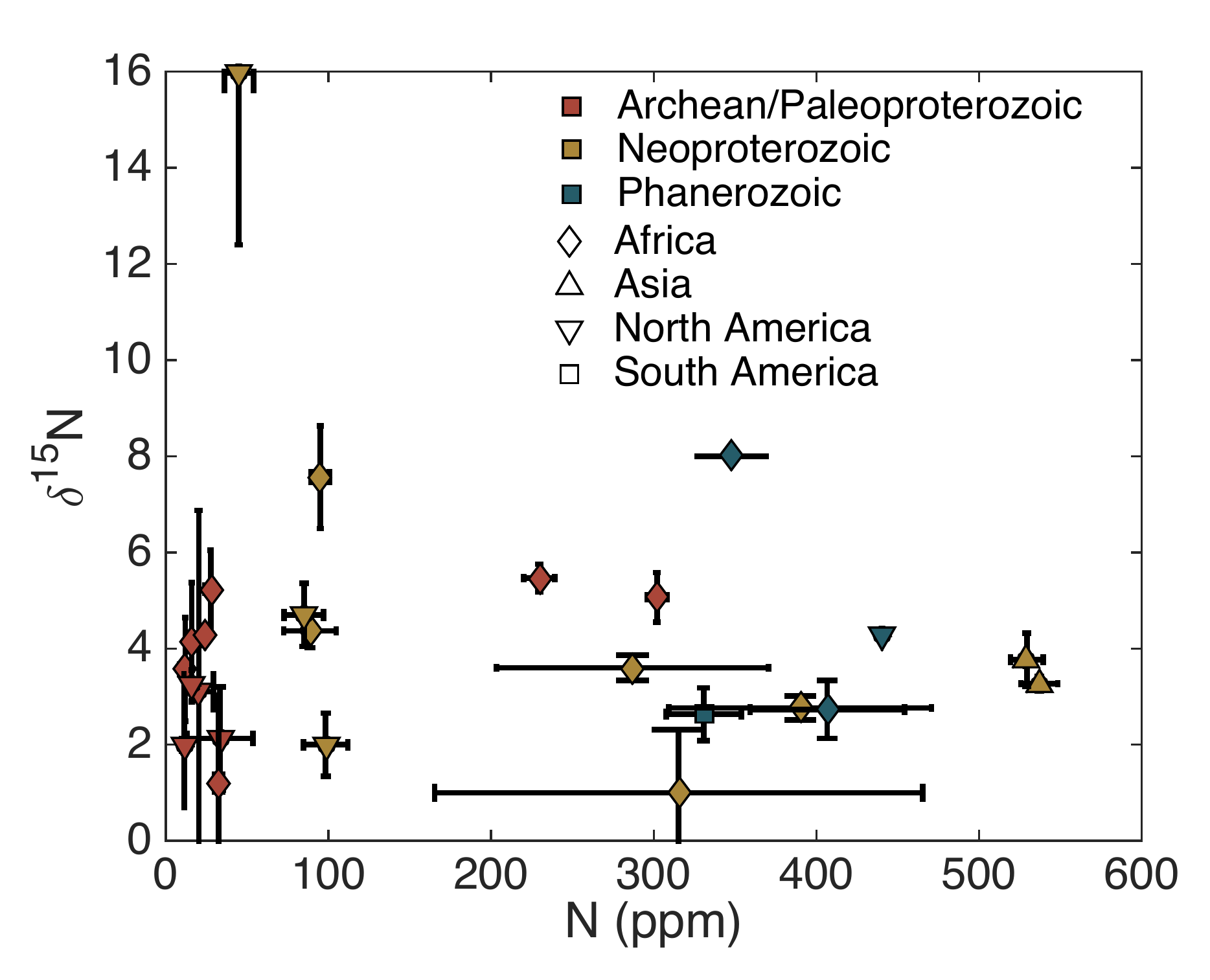}
\caption{Nitrogen isotope values plotted against N concentration. Lack of correlation between isotopes and concentrations from samples of all ages suggests that there has not been N loss during diagenesis. Nitrogen loss tends to result in samples with low N concentration having high $\delta^{15}$N values, which is not observed. }
\label{fig:del15N_N}
\end{figure*}

Crucial to our presented interpretation is demonstrating that the N concentrations have not been altered through time. That is, the lower concentrations observed in older rocks are not simply the result of progressive N loss through time. There are a number of approaches to assess this possibility. Firstly, we observe no correlation between $\delta^{15}$N values and N concentration (Fig. \ref{fig:del15N_N}). If progressive N loss was occurring, the expected trend would be higher $\delta^{15}$N values associated with lower N concentrations, due to the preferential loss of $^{14}$N during diagenesis. 

\begin{figure*}[]
\includegraphics[width=\textwidth]{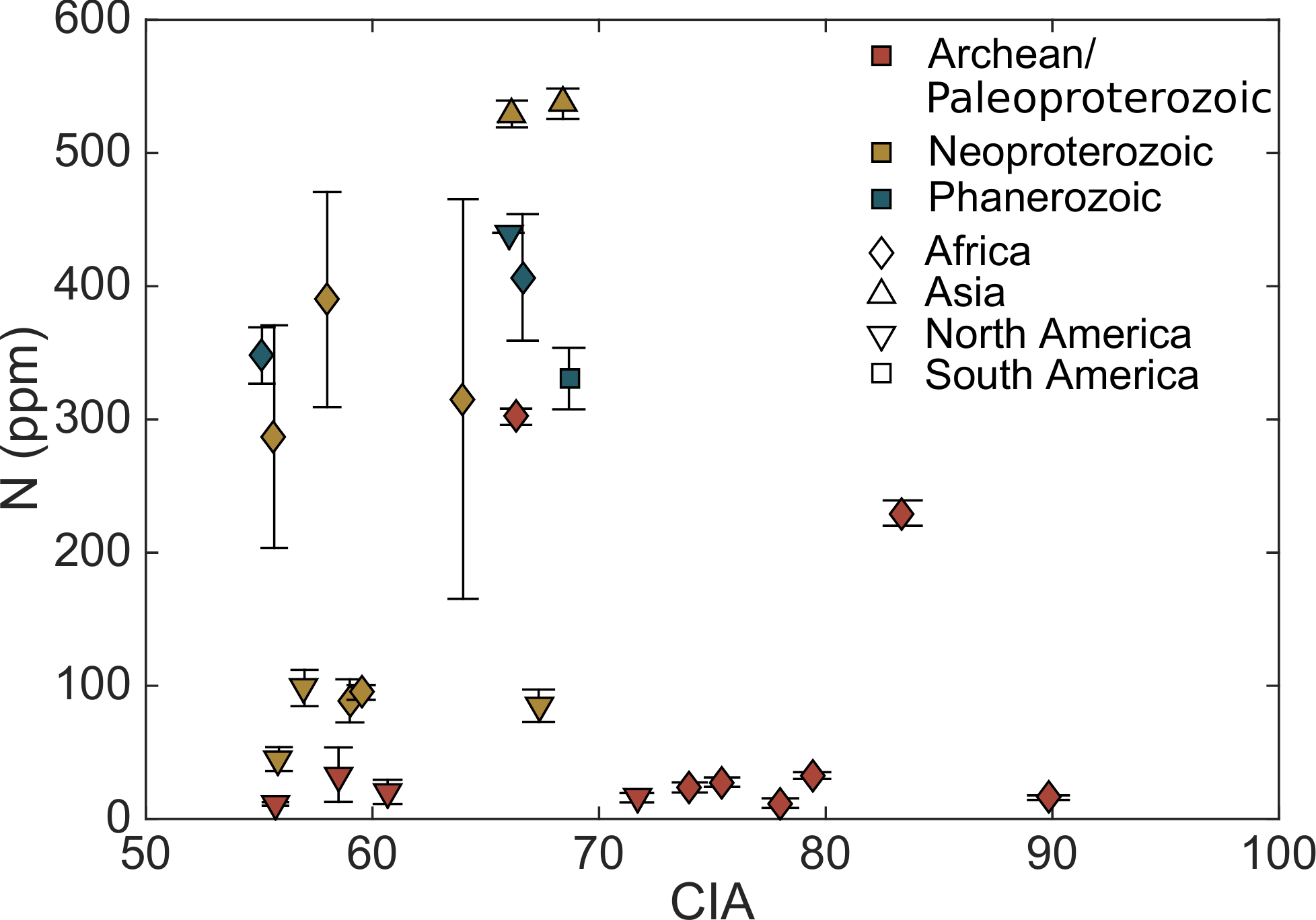}
\caption{Nitrogen concentration plotted against Chemical Index of Alteration (CIA). See text for details, but lack of correlation suggests that aqueous alteration alone cannot explain the increase in N concentration through time. }
\label{fig:CIA}
\end{figure*}

Secondly, comparisons with other elemental compositions indicate lack of alteration through time. As discussed in detail in \citep{Gaschnig_et_al_2016}, a first-pass approach is to use the Chemical Index of Alteration (CIA), which is defined as: \ce{Al2O3}/(\ce{Al2O3} + \ce{CaO}* + \ce{Na2O} + \ce{K2O}) $\times100$ \citep{Nesbitt_and_Young_1982}. This index supposes that alteration of feldspars to clay minerals during chemical alteration and weathering will increase the CIA. There is no observed correlation between N concentration and CIA in these samples (Fig. \ref{fig:CIA}). As discussed in \citet{Gaschnig_et_al_2016}, till samples with a high CIA likely inherited this signal from a weathered source rock, rather than having experienced extensive chemical weathering themselves. 

\begin{figure*}[]
\includegraphics[width=\textwidth]{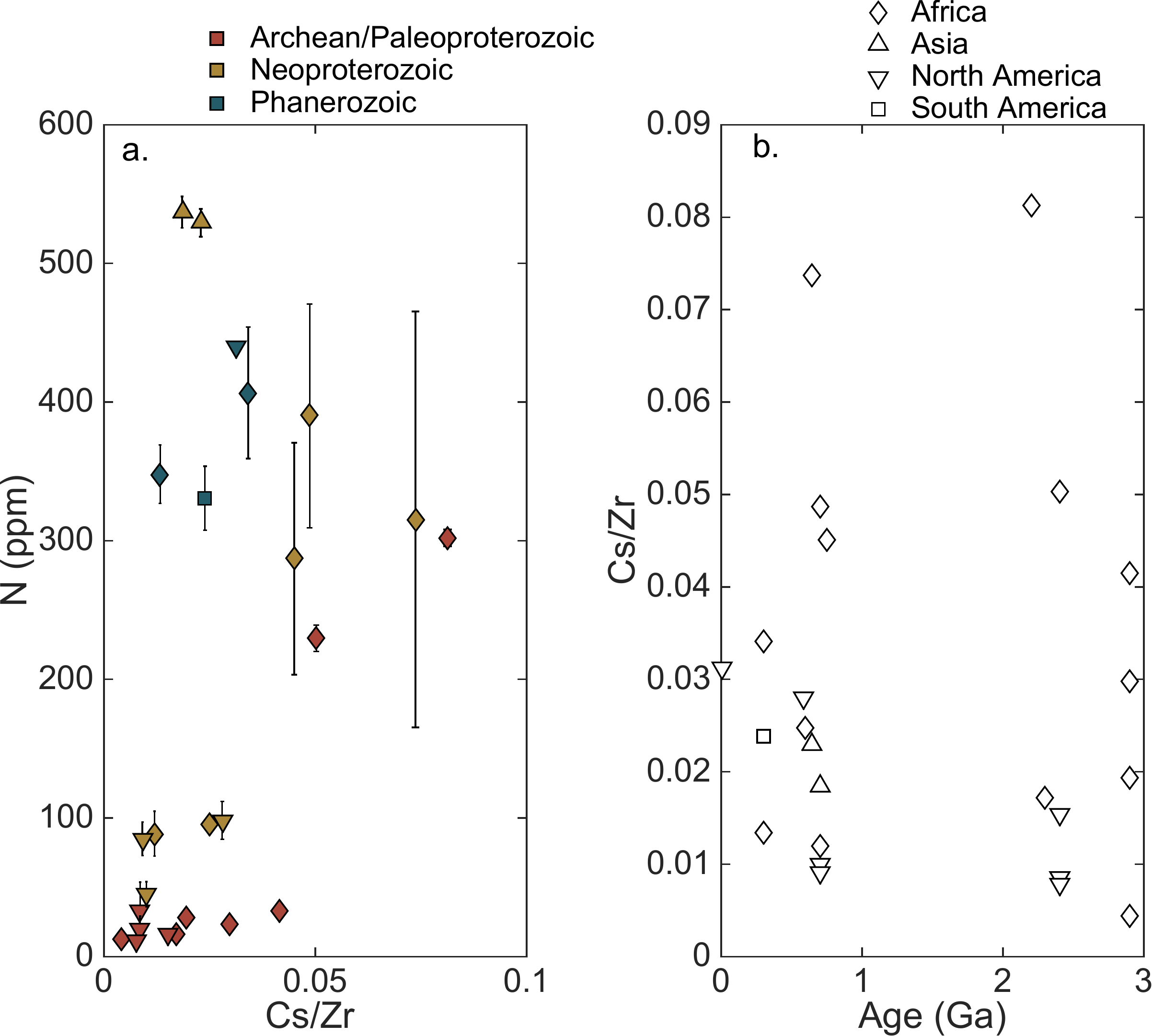}
\caption{Nitrogen concentration plotted against Cs/Zr. See text for details, but lack of correlation suggests that aqueous alteration alone cannot explain the increase in N concentration through time, with the possible exception of the lowest N samples. There is no change in the Cs/Zr ratio through time, suggesting that geographic and temporal evolution of Cs and Zr does not explain variation seen in a.}
\label{fig:Cs/Zr}
\end{figure*}

An additional comparison can be made using Large Ion Lithophile (LILE) and High Field Strength Elements (HFSE). Both these groups are incompatible, but in general LILE are fluid-mobile and HFSE are not. We use Cs to represent LILE and Zr to represent HFSE.  Neither Zr nor Cs abundances in the crust have shown secular changes through time \citep{Gaschnig_et_al_2016}. If progressive N loss via aqueous alteration through time was the sole cause for the trend in increased N through time, samples that have low N would have a correspondingly low Cs/Zr. While samples with the very lowest N and Cs/Zr are from the Archean/Palaeoproterozoic (Fig. \ref{fig:Cs/Zr}), there are also a number of younger samples with low Cs/Zr and high N.Thus, we suggest that post-depositional aqueous alteration alone cannot explain the trend of decreased N concentrations back in time, though we cannot rule out some alteration for the lowest N samples.

\begin{figure*}[]
\includegraphics[width=\textwidth]{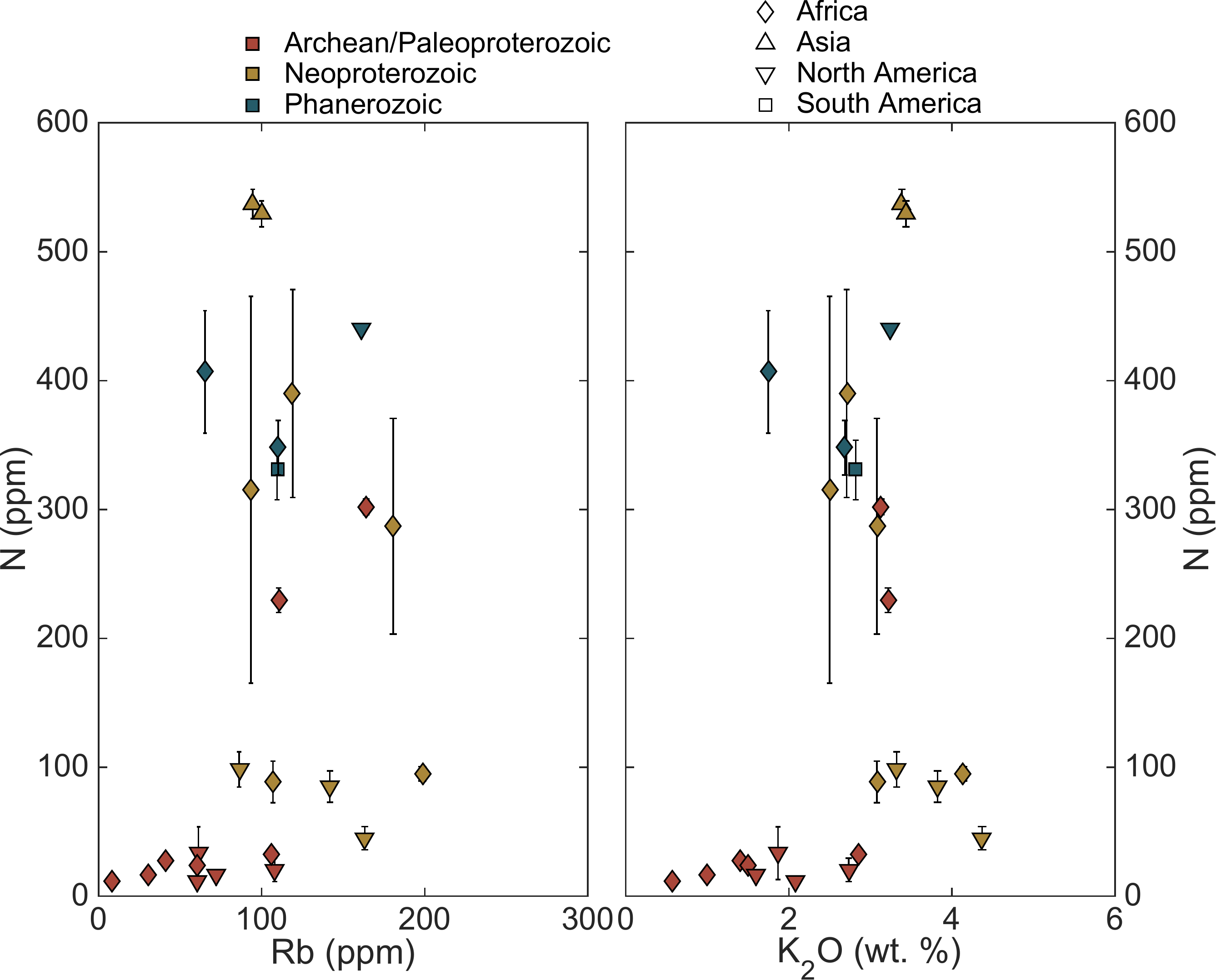}
\caption{Nitrogen concentration plotted against Rb and \ce{K2O}. We note that for low N samples there is a positive correlation, but for moderate and high N samples, no correlation is observed. }
\label{fig:N_Rb_K}
\end{figure*}

\section*{Mineral hosts for N} 

\begin{figure*}[]
\includegraphics[width=\textwidth]{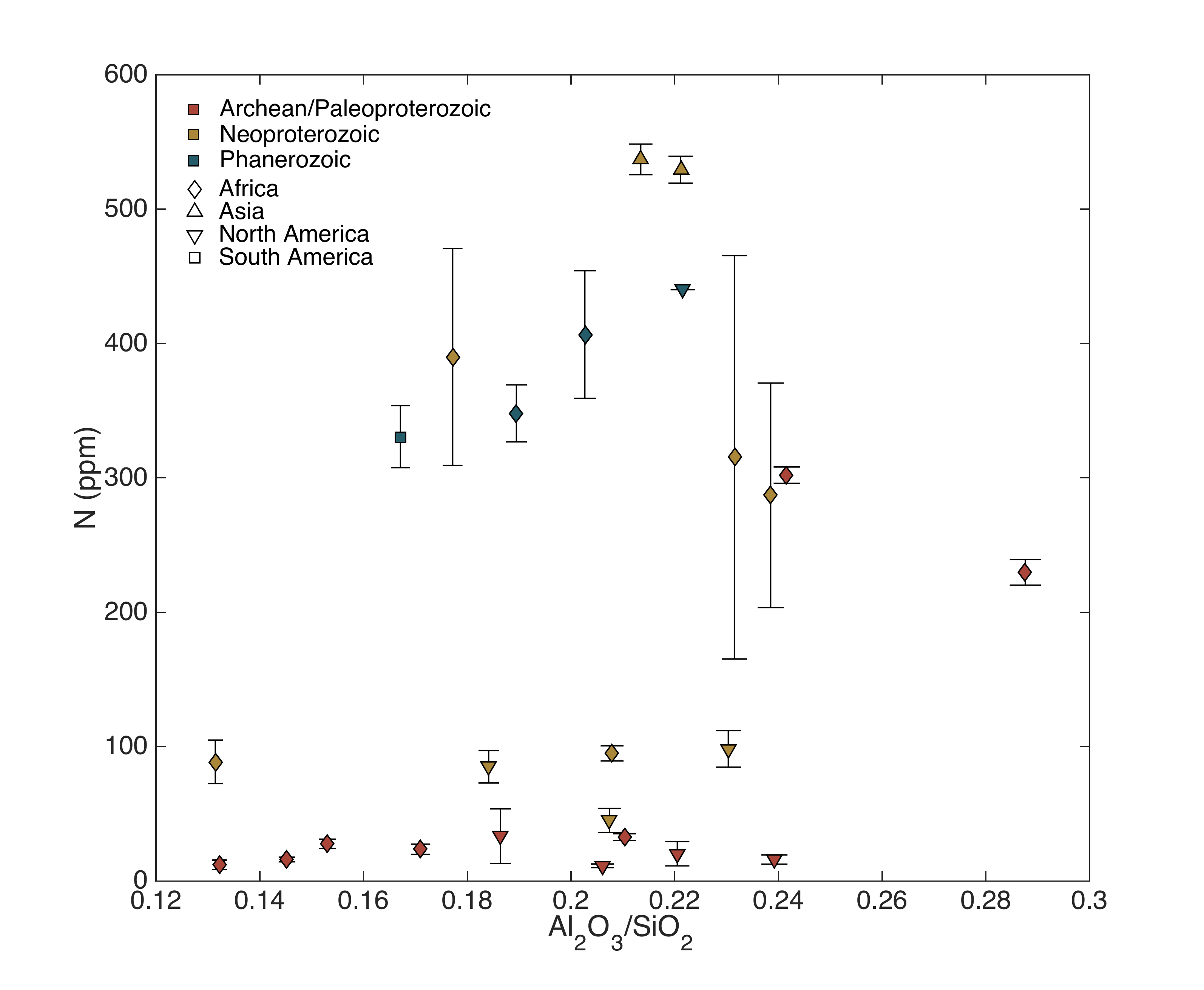}
\caption{Nitrogen concentration plotted against \ce{Al2O3}/\ce{SiO2}. }
\label{fig:N_AlSi}
\end{figure*}

We do not know the exact mineral host of N in these samples. Typically, \ce{NH4+} is the most common geological species of N, though there may be small amounts of organic N as well. In detail, some samples that have high N also have a high \ce{Al2O3}/\ce{SiO2} ratio (Fig. \ref{fig:N_AlSi}), which is indicative of a weathering environment rich in feldspars and clays compared to quartz \citep[][and references therein]{Gaschnig_et_al_2016}. Thus, for many samples, K-bearing phases are a likely host for N as \ce{NH4+}, though the lack of correlation between N and Rb (which also substitutes for K), indicates that this simple relationship may not be true for all samples (Fig. \ref{fig:N_Rb_K}).

Palaeoproterozoic samples from South Africa exemplify this relationship (Fig. \ref{fig:N_AlSi}). Other high N samples, such as those from North America, have a similar \ce{Al2O3}/\ce{SiO2} ratio as higher N samples from Africa and South America, indicating clays/feldspars may not be the main host for N. Another mineral may be the host in these settings. 

\section*{Nitrogen concentrations by continent}

\begin{figure*}[]
\includegraphics[width=\textwidth]{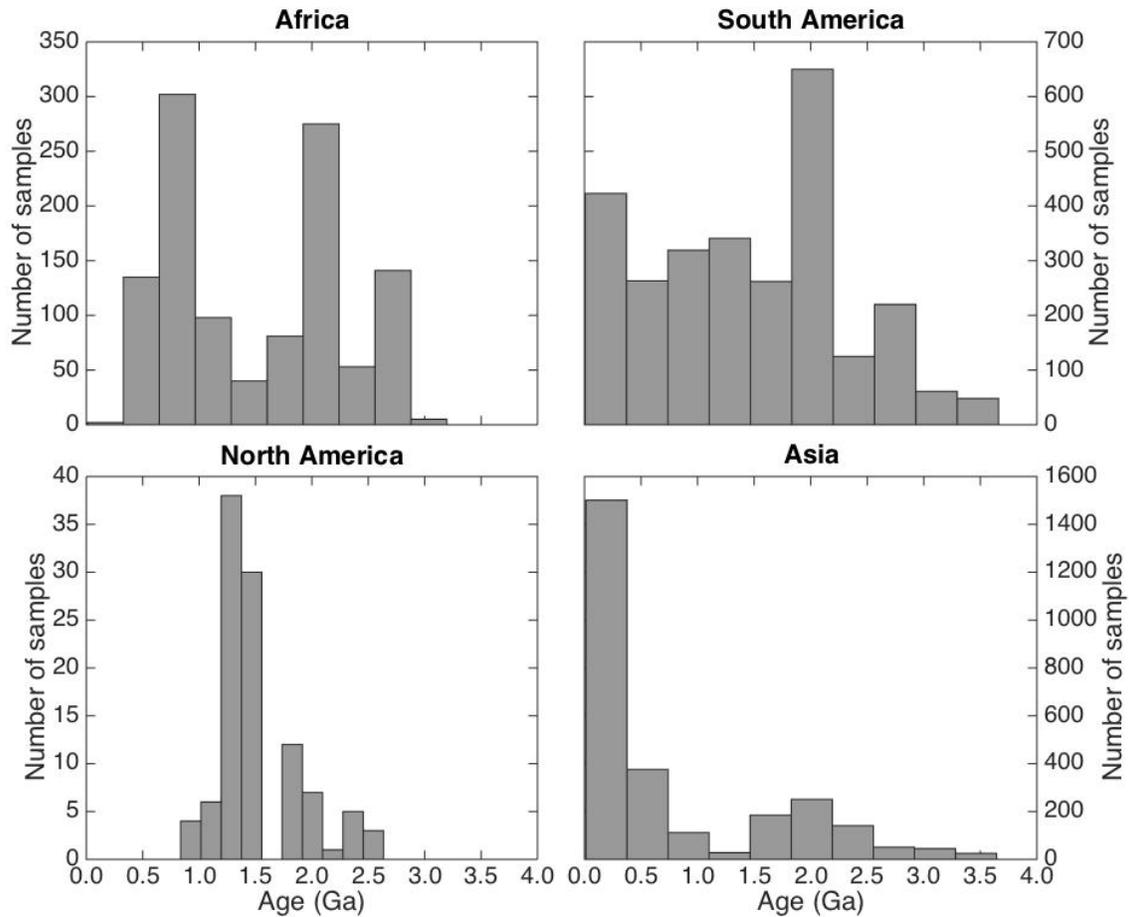}
\caption{Histograms of detrital zircons from Africa, South America, North America, and Asia through time. African zircons have two peaks at 2.1 and 0.75 Ga, while South America has the peak at 2.1 Ga, and perhaps a peak from 0 to 0.5 Ga. Both North America and Asia have a single dominant peak, at 1.2 Ga and 0-0.5 Ga, respectively. It is possible that peaks in ages correspond to periods of continental growth and, by extension, periods of N-sequestration in the crust.}
\label{fig:zircons}
\end{figure*}

As mentioned in the main text, there is an apparent relationship between the current continent of each sample and  N concentration. That is, samples from Africa, South America, and Asia are generally high while samples from North American are generally low. The strongest correlation both regionally and globally is age, but another possibility that could explain some of the geographic control is a different history of continental growth and assemblage. Zircon ages from Africa and North America from the compilation of \citep{Belousova_et_al_2010} show two peaks and one peak in ages, respectively (Figs. \ref{fig:zircons}). 

 If these zircon ages peaks correspond with periods of enhanced continental growth, it would suggest that Africa grew somewhat earlier than North America, and correspondingly biologically processed N was incorporated during this phase of continental growth. A major continental growth period occurred later in North America, around 1.2 Ga, with an increase in N in till samples not seen till the Phanerozoic. There would be some lag time between continental growth and erosion by glaciers, thus these periods of growth represent a maximum hypothesized age of N incorporation for each continent. A reasonable test of this pulsed N addition hypothesis would be the analysis of N concentration in felsic intrusive igneous rocks, which are more temporally associated with the continental-growth phases. Though there are few till samples from Asia and South America, distinct zircon age peak distributions indicate possible different continental growth histories, with correspondingly distinct N histories.

It is also possible that paleogeography could exert a control over timing of N incorporation. However, if the source of N to the continental crust is biologically processed, subduction zone transported material, both the manner of biologic processing and geometry and extent of subduction zones should have a greater effect. Perhaps, though, if continental subduction zones are located nearer to more productive areas, more N could be buried with organic matter and processed in a subduction zone. Indeed, areas with high productivity near Central America have more N in sediments than low-productivity areas in the eastern Pacific \citep{Elkins_et_al_2006, Mitchell_et_al_2010}. 

In the modern ocean, areas with high levels of N-fixing (driving atmospheric N2 drawdown) are mostly identified in the tropics. Thus, it is possible that high N-fixing or high productivity areas overlying subduction zones could enhance incorporation of N into continental crust. The distribution of N-fixing regions throughout geologic time are poorly constrained; we are not aware of any data speaking to this directly. In addition, the temperature/pH/redox of subduction zones seems to exert strong control over the fate of subducted N \citep{Busigny_et_al_2011, Mikhail_and_Sverjensky_2014}, and might be more important than paleogeography.

\begin{figure*}[]
\includegraphics[width=\textwidth]{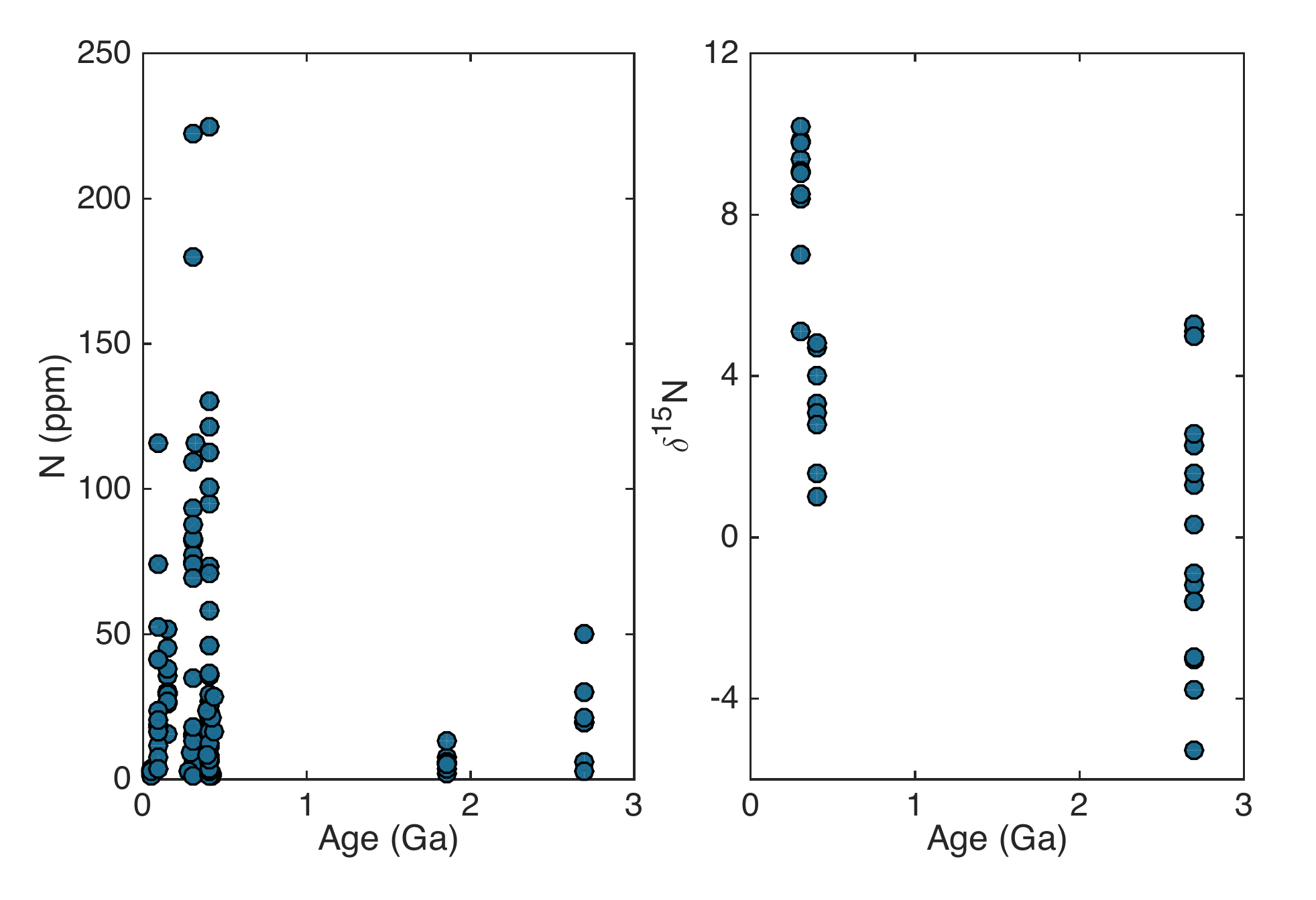}
\caption{Nitrogen concentration and $\delta^{15}$N values of granitic rocks through time. These data are consistent with an increase in N content of continental crust through time, with increase in isotopic values consistent with a biologically processed source for said N.  }
\label{fig:granitic}
\end{figure*}

\section*{Granitic N}

Nitrogen concentration and isotopic data from the compilation of \citet{Johnson_and_Goldblatt_2015} are shown in Fig. \ref{fig:granitic}. These are data from both whole rock and mineral specific N analyses, scaled up to whole rock values. The rocks analysed are from the British Isles \citep{Hall_1987, Hall_1988, Hall_1993, Cooper_and_Bradley_1990, Bebout_et_al_1999a}, southern Europe \citep{Hall_et_al_1991,Hall_1999}, the Canadian shield \citep{Jia_and_Kerrich_2000}, Iran \citep{Ahadnejad_et_al_2011}, Finland \citep{Itihara_and_Suwa_1985}, and the United States \citep{Hoering_1955}. While data are sparse, especially isotopic data, they are consistent with an increase in crustal N through time. Samples from the Archean and Palaeoproterozoic have a mean N concentration of 16 ppm, while those from the Phanerozoic average 43 ppm N. These means are statistically different as shown by  Student's t-test \citep{Student_1908}. Isotopic data may increase from depleted, mantle-like values towards more enriched, biologic/sedimentary values through time. Analyses are heavily biased towards Europe, and especially the British Isles. Additional geographic coverage, in addition to temporal coverage, would greatly help with future interpretations.

\setstretch{1.15}
{\vskip 0.1cm}
\bibliographystyle{apalike}
%\bibliography{\string~/Science/references/refs.bib}

\end{document}